\documentclass{article}
\usepackage{amssymb}
\usepackage{amsmath} 
\usepackage{epsfig}
\begin{document}
\numberwithin{equation}{section}
\title{\bf Solution of the Dirichlet boundary value problem
for the Sine-Gordon equation\footnote{Work supported in part by
by INTAS 99-1782}}
\author{J\'er\^ome LEON \\ 
{\em Physique Math\'ematique et Th\'eorique, CNRS-UMR5825,}\\ 
Universit\'e Montpellier 2, 34095 MONTPELLIER (France)} \date{}\maketitle

\begin{abstract} The sine-Gordon equation in light cone coordinates is solved
when  Dirichlet conditions on the L-shape boundaries of the strip
$\{t\in[0,T]\}\cup\{x\in[0,\infty]\}$ are prescribed in a class of functions
that vanish (mod $2\pi$) as $x\to\infty$ at initial time.  The method is based
on the inverse spectral transform (IST) for the Schr\"odinger spectral problem
on the semi-line $x>0$ solved as a Hilbert boundary value problem. Contrarily
to what occurs when using the Zakharov-Shabat eigenvalue problem, the spectral
transform is regular and in particular the discrete spectrum contains a finite
number of eigenvalues (and no accumulation point). \end{abstract}

\section{Introduction}
The sine-Gordon equation in light cone coordinates (dimensionless variables,
subscript stands for partial derivative) for the field $\theta(x,t)$,
\begin{equation}\label{sg}
\theta_{xt}+\sin\theta=0\ ,\end{equation}
is one of  the most widely studied nonlinear wave equation because of its 
intrinsic mathematical  properties and its wide applicability in physics. 
The initial value problem on the infinite line $x\in(-\infty,+\infty)$, namely
the initial datum $\theta(x,0)$ together with the asymptotic behaviors
$\theta=2n\pi$  as $x\to\pm\infty$, has been solved by the {\em inverse
spectral transform} (IST) in \cite{ist-sg,takh-fad}.

The boundary value problem for \eqref{sg} on the quadrant $\{x>0,\ t>0\}$
has been solved in \cite{fokas} and reconsidered in \cite{alex}. In both
approaches it has been shown that a constant boundary datum $\theta(0,t)$
on an initial vacuum $\theta(x,0)=0$ produces an infinite series of discrete
eigenvalues with accumulation point at the origin. The problem is then that the
Jost solutions are not continuous in a neighborhood of the real axis and the
spectral theory fails.

We solve this problem by considering an alternative Lax pair for the
sine-Gordon equation deduced from the work \cite{marflo} on two-dimensional
extensions of sine-Gordon. The principal spectral operator is the Schr\"odinger
eigenvalue problem with {\em complex potential} that we reconsider on the
half-line $x>0$ with a given boundary value of the potential in $x=0$.

The spectral theory developped here follows the approach introduced in
\cite{sasha} for stimulated Raman scattering, reformulated in \cite{alex}, and
later applied to others continuous \cite{jl-fx} or discrete \cite{jl-boit}
coupled wave systems. In the next section the Lax pair is set and  the boundary
value problem is discussed.  Section \ref{sec:direct} is devoted to the
solution of the {\em direct spectral problem}, namely to the method of
construction of the spectral transform, or {\em nonlinear Fourier transform},
of the sine-Gordon field.

The time dependence of the spectral transform is then derived in section
\ref{sec:sg} where it appears to be a nonlinear Riccati type evolution.  It is
examined for particular boundary data and shown to lead asymptotically to a
finite set of discrete eigenvalues (soliton component of the spectrum).  In
section \ref{sec:invpro} we solve the {\em inverse problem} which is the
reconstruction of the sine-Gordon field from the spectral transform. This is
done by formulating  the reconstruction as a Hilbert boundary value
problem.  In doing this attention must be paid to the asymptotic boundary
values in the spectral plane required for the solutions of this Hilbert
boundary value problem. Indeed one of the Jost solution has an essential
singularity for large real values of the spectral parameter. Still the solution
can be obtained by a Cauchy-Green integral equation with a contour of
integration that differs from the one used in the infinite line case.  The
spectral theory of the Schr\"odinger operator on the half-line can also be
developped within the Gel'Fand-Levitan-Marchenko fromalism as shown in
\cite{pierre} where the boundary value problem for the Korteveg-de Vries
equation is studied.

\section{Alternative Lax pair for sine-Gordon}

\paragraph*{Weakly commuting operators.}

Let us consider the following pair of scalar operators: 
the {\em principal spectral operator} 
\begin{equation}\label{lax-x}
 T_1=\partial_{xx}+k^2-p\ ,\end{equation}
where $p(x,t)$ is the {\em principal potential}, and the
{\em auxiliary spectral operator} 
\begin{equation}\label{lax-t}
T_2=\partial_{xt}+\alpha\partial_t+\beta \ ,\end{equation}
where $\alpha(x,t)$ and $\beta(x,t)$ are the {\em auxiliary potentials}.

Originaly introduced for two-dimensional problems \cite{flomar}, the {\em weak
compatibility} of those operators results in requiring that they commute on
their common eigenspace.  In other words
\begin{equation}
[T_1,T_2]=f\ T_1 +g\ T_2
\ ,\end{equation}
where $f$ and $g$ are two unknown functions. Straightforward examination of
highest order derivatives gives
\begin{equation}
f=0\ ,\quad g=2\alpha_x\ ,\end{equation} 
hence we are left with the operatorial equation
\begin{equation}\label{triplet}
[T_1,T_2]=2\alpha_xT_2 \ .\end{equation}
This equation  exlpicitely implies the following system
\begin{align}
&p_x=-\alpha_{xx}+2\alpha\alpha_x\ ,\label{weak1}\\
&p_t=-2\beta_x\ ,\label{weak2}\\
&p_{xt}+\alpha p_t=-\beta_{xx}+2\alpha_x\beta\ ,\label{weak3}
\end{align}
in which two equations allow to eliminate two fields and the remaining 
equation is the nonlinear evolution.

\paragraph*{The sine-Gordon equation.}

Assuming for all $t$ the asymptotic behaviors (we will prove later that
these behaviors {\em result} from the properties of the initial datum)
\begin{equation}\label{asymp-x}
x\to\infty\ :\ \left\{p\to 0\ ,\ \alpha\to0\ ,\ \alpha_x\to0
\ ,\ \beta_x\to0\right\}\ ,\end{equation}
two of the above equation can be integrated and give 
\begin{equation}\label{weak}
\beta_x=-2\alpha\beta\ ,\quad p=\alpha^2-\alpha_x\ ,
\end{equation}
while the remaining equation reads as the evolution
\begin{equation}p_t=-2\beta_x\ .\end{equation}
To eliminate two fields we define the new variable $\theta(x,t)$ by
\begin{equation}
\beta=\frac14e^{i\theta}\ ,\end{equation}
which from \eqref{asymp-x} will be assumed to obey
\begin{equation}\label{theta-x}
\theta_x\underset{x\to\infty}{\longrightarrow}0\ .\end{equation}
The relation \eqref{weak} then gives $p$ and $\alpha$  in terms of $\theta$ as
\begin{equation}\label{p-alpha-theta}
\alpha=-\frac i2\theta_x\ ,\quad p=\frac i2\theta_{xx}-\frac14\theta_x^2
\ ,\end{equation}
where one can check that the behavior \eqref{theta-x} ensures the asymptotic 
expressions \eqref{asymp-x}.

The evolution $p_t=-2\beta_x$  can then be factorized as
\begin{equation}\{\frac\partial{\partial x}\ +i\theta_x\}
\left(\theta_{xt}+\sin\theta\right)=0\ .\end{equation}
This can be integrated once and the integration constant can be evaluated
by reading the evolution as $x\to\infty$ where $\theta\to\lambda$. The results
is the following complex extension of the sine-Gordon equation
\begin{equation}\label{sg-complex}
\theta_{xt}+e^{i\lambda}\sin(\theta-\lambda)=0\ ,\quad
\lambda=\lim_{x\to\infty}\theta
\end{equation}
associated to the asymptotic behavior \eqref{theta-x}. Note that in general
the parameter $\lambda$ is a function of time $t$.

In the case of real valued $\theta$, and hence real valued $\lambda$, 
the above evolution implies $\lambda=0$ and hence reduces to 
\begin{equation}\label{sg-real}
\theta\in{\mathbb R}\ \Rightarrow\ \theta_{xt}+\sin\theta=0\ ,\quad
\theta\underset{x\to\infty}{\longrightarrow}0\mod 2\pi
\ .\end{equation}
Thus the standard sine-Gordon equation results as the {\em weak compatibility}
of operators $T_1$ and $T_2$ for real valued field $\theta$. 

It is important to remark here that the imposed vanishing asymptotic behavior
of the derivative $\theta_x$ {\em for all $t$} as given in \eqref{theta-x}
actually results from the above assumed asymptotic behavior of $\theta$ as soon
as $\theta_x$ vanishes at infinity at initial time.  Indeed, by integration of
the sine-Gordon equation we can write
\begin{equation}\label{integ-sg}
\theta_x(x,t)=\theta_x(x,0)-\int_0^t d\tau\ \sin\theta(x,\tau)\ .
\end{equation}
This expression serves actually as a means to go from Dirichlet to Neuman
conditions on the boundaries $x=0$ and $x=\infty$.
It is also useful to prove that the assumed vanishing asymptotic behavior
of the principal potential $p(x,t)$ results from the asymptotic value of
$\theta(x,t)$. Indeed, upon derivation of \eqref{integ-sg} it is easy to
obtain that if the initial datum is such that $\theta_x(x,0)$ and
$\theta_{xx}(x,0)$ vanish as $x\to\infty$, then $\theta_{xx}$ vanishes for
all time as $x\to\infty$.

Note also that the expression \eqref{sg-complex} of the sine-Gordon equation
for complex variable can be put in the form \eqref{sg-real} by the following
change of variables
\begin{equation}\label{sg-varphi}
\varphi=\theta-\lambda\ ,\quad \tau=\int_0^tdt'e^{-i\lambda(t')}\ ,
\end{equation}
which gives
\begin{equation}\varphi_{x\tau}+\sin\varphi=0\ ,\quad\varphi
\underset{x\to\infty}{\longrightarrow}0\mod 2\pi
\ .\end{equation}
We have assumed hereafter the general situation where the constant $\lambda$
is allowed to depend on time $t$.

\paragraph*{Dirichlet boundary data.}

We consider then the sine-Gordon equation \eqref{sg}, a priori for a complex 
field  $\theta$, on the strip
\begin{equation}
\{x\in [0,\infty]\}\cup\{t\in[0,T]\}\ ,\end{equation}
for any value of $T$. The Dirichlet conditions are the data of the
function $\theta(x,t)$ on the L-shape boundaries of this strip, namely
\begin{equation}\label{dirichlet}
t=0\ :\ \theta(x,0)=\tilde\theta(x)\ ,\quad x=0\ :\ \theta(0,t)=\theta_0(t)\ .
\end{equation}

The auxiliary spectral operator \eqref{lax-t} explicitely contains the
variable $\alpha(x,t)$ which has then to be evaluated in $x=0$ explicitely
in terms of the Dirichlet data \eqref{dirichlet}. By expressing 
\eqref{integ-sg} in $x=0$ we have
\begin{equation}\label{theta-x-0}
\theta_x(0,t)=\tilde\theta_x(0)-\int_0^td\tau\ \sin\theta_0(\tau)\ ,
\end{equation}
which from \eqref{p-alpha-theta} gives
\begin{equation}\label{alpha-0}
\alpha_0(t)=\frac i2\int_0^td\tau\ \sin\theta_0(\tau)-
\frac i2\tilde\theta_x(0)\ ,
\end{equation}
where $\alpha_0=\alpha(0,t)$ is then completely determined from 
\eqref{dirichlet}. Similarly we have to evaluate $p_0(t)=p(0,t)$ occuring
in the principal Lax operator. By derivation of \eqref{integ-sg} we
obtain
\begin{equation}\label{p-0}
p_0(t)=\frac i2\tilde\theta_{xx}(0)-\frac i2\int_0^td\tau\ \theta_x(0,\tau)
\cos\theta_0(\tau)-\frac14\theta_x^2(0,t)\ ,
\end{equation}
where $\theta_x(0,t)$ is explicitely given in \eqref{theta-x-0}. We have then
obtained that the data of $\theta_0(t)$ and $\tilde\theta(x)$ determine
completely the boundary values in $x=0$ of the {\em potentials} of the 
fundamental operators $T_1$ and $T_2$.

For the other boundary $x=\infty$ we first select a class of functions for the
initial datum $\tilde\theta(x)$. As usual for implementation of the spectral
transform, let us assume that $\tilde\theta(x)$ belongs to the Schwartz space
$\cal S$ (bounded functions that vanish with all their derivatives faster than
any power of $1/x$). Although this condition is rather stringent, it fits well
a generic physical situation, in particular the case of an initial {\em
vacuum}. It is then easy to obtain from \eqref{integ-sg} the property
\begin{equation}\label{th-asymp}
\left\{\tilde\theta(x)\in{\cal S}\ ,\quad 
\theta(\infty,t)=0\mod 2\pi\right\}
\ \Rightarrow\ \theta_x(\infty,t)=0\ . \end{equation}
This can be worked again for the second derivative and finally the potentials
$p(x,t)$ and $\alpha(x,t)$ do vanish as $x\to\infty$ and $\beta$ goes
asymptotically to $1/4$.

\section{Spectral problem\label{sec:direct}}

The Schr\"odinger spectral problem on the half-line $T_1\psi=0$ with complex 
potential, namely
\begin{equation}\label{spectral-psi}
x>0\  ,\quad \psi_{xx}+(k^2-p)\psi=0\  ,\quad p(0)=p_0\ ,\end{equation}
is solved now and the inverse problem will be takled by means of a Hilbert
boundary value problem. A partial solution has already been given in
\cite{jl-fx}, in the case of real valued potential. 
We rework everything here with slightly different definitions
and in particular give the complete solution of the Hilbert boundary
value problem. Note that this spectral problem has also been considered
in \cite{thana-bea}.

\paragraph*{Basic solutions.}

 We redefine the spectral equation as
\begin{equation}\label{spectral-phi}\psi=\phi\ e^{ikx}\ ,\quad
\phi_{xx}+2ik\phi_x-p\phi=0\  ,\end{equation}
and select two basic solutions $\phi^+$ and $\phi^-$ by
\begin{align}
\phi^-(k,x)=1&+\frac1{ 2ik}\int_{0}^xdy\ \phi^-(k,y)p(y)
[1-e^{-2ik(x-y)}]\ ,\label{phi-}\\
\phi^+(k,x)=1&+\frac1{2ik}\int_{0}^xdy\ \phi^+(k,y)p(y)
\notag\\
&+\frac1{2ik}\int_x^{\infty}dy\ \phi^+(k,y)p(y)e^{-2ik(x-y)}\ .
\label{phi+}\end{align}
where conditions on $p(x)$ are assumed (e.g. $p(x)$ belongs to the Schwartz
space) that ensure uniqueness of the solution of the above integral equations.  It
is useful  to compute the derivatives
\begin{align}
&\phi^-_x(k,x)=\int_{0}^xdy\ \phi^-(k,y)p(y)
e^{-2ik(x-y)}\ ,\label{phi-x}\\
&\phi^+_x(k,x)=-\int_x^{\infty}dy\ \phi^+(k,y)p(y)e^{-2ik(x-y)}\ .
\label{phi+x}\end{align}
An important property for the following is that
another set of basic solutions is constituted by $\phi^\pm(-k,x)e^{-2ikx}$.
Another important property is the fact that the wronskian of two solutions
of \eqref{spectral-psi} ia a constant, which means for any two solutions
$\phi_1$ and $\phi_2$ of \eqref{spectral-phi}
\begin{equation}\label{wronsk}
W(\phi_1,\phi_2)=\phi_{1,x}\phi_2-\phi_1\phi_{2,x}=C\ e^{-2ikx}\ ,
\end{equation}
where $C$ is a constant.

\paragraph*{Spectral transform.}

The solution $\phi^+$ is called the {\em physical solution} and serves to
define the scattering coefficients as follows.
The boundary value in $x=0$ 
\begin{equation}\label{phi+zero}
x=0\ :\  \phi^+= 1+\rho(k)\ ,
\end{equation}
defines the {\em reflection coefficient} $\rho(k)$ as
\begin{equation}\label{rho-def}
\rho(k)=\frac{1}{2ik}\int_{0}^{\infty}dy\ p(y)\phi^+(k,y)e^{2ik y}\ ,
\end{equation}
and consequently \eqref{phi+x} gives
\begin{equation}\label{phi_x+zero}
x=0\ :\  \phi^+_x= -2ik\rho(k)\ .
\end{equation}
The boundary value in $x=\infty$ 
\begin{equation}\label{phi+infty}
x\to\infty\ :\  \phi^+\to \tau(k)\ ,
\end{equation}
defines the {\em transmission coefficient} $\tau(k)$ as
\begin{equation}\label{tau-def}
\tau(k)=1+\frac{1}{2ik}\int_{0}^{\infty}dy\ p(y)\phi^+(k,y)\ ,
\end{equation}
and we have also
\begin{equation}\label{phi_x+infty}
x\to\infty\ :\  \phi^+_x\to 0\ .
\end{equation}
The wronskian relation \eqref{wronsk} applied to the solutions $\phi^+(k)$
and $\phi^+(-k)e^{-2ikx}$ at $x=0$ and $x=\infty$ successively gives
\begin{equation}\label{unitarity}
\rho(k)\rho(-k)+\tau(k)\tau(-k)=1\ ,\end{equation}
called the {\em unitarity relation}. 

 From its very definition, the spectral tranfrom $\rho(k)$ seems to be ill 
defined in $k=0$. Actually we recall in the appendix \ref{app:k=0} that
\begin{equation}
k=0\ :\ \rho=-1\ ,\quad \tau=0\ ,\end{equation}
which is well known in the infinite line case \cite{sabatier}.

\paragraph*{Hilbert boundary value problem}

The solution $\phi^-$ is called the {\em intermediate solution} and serves to
complete the basis. We can for instance express $\phi^+(k)$  in
terms of the basis $\phi^-(k)$ and $\phi^-(-k)e^{-2ikx}$, the coefficients
being computed by comparing the values in $x=0$ of both the functions and
their derivatives. We obtain
\begin{equation}\label{hilbert}
\phi^+(k,x)-\phi^-(k,x)=\rho(k)e^{-2ikx}\phi^-(-k,x)\ ,\end{equation}
which constitutes a Hilbert problem of the real $k$-axis, we shall come
to that argument later. For the moment we express the above relation by
changing $k$ to $-k$ and solve the resulting linear system for the
function $\phi^-(k)$ to obtain
\begin{equation}
\phi^-(k,x)=\frac1{1-\rho(k)\rho(-k)}\left[
\phi^+(k,x)-\rho(k)e^{-2ikx}\phi^+(-k,x)\right]\ .\end{equation}
Using the definitions of $\phi^-$ and $\phi^-_x$ and the above relation we can
compute the different boundary values and get
\begin{align}
x=0\quad :&\ \phi^-= 1\ ,\quad  \phi^-_x=0\ ,\label{phi-0}\\
x=\infty\quad :&\ \phi^-\sim\frac1{\tau(-k)}-\frac{\rho(k)}{\tau(k)}e^{-2ikx}
\ ,\quad  \phi^-_x\sim
2ik\frac{\rho(k)}{\tau(k)}e^{-2ikx}
 \ ,\label{phi-infty}
\end{align}
where we have also used the unitarity relation \eqref{unitarity}.

\section{Solution of sine-Gordon\label{sec:sg}}

We have constructed in the preceding section the spectral transform $\rho(k)$
of the principal potential $p(x)$. We shall prove in section \ref{sec:invpro}
that the datum of $\rho(k)$ in the upper half plane completely determines
the potential $p(x)$, and we shall give the tools to reconstruct this potential.
But we consider first the question of the {\em time dependence} of the spectral
transform when the potential depends on time according to the evolutions
\eqref{weak}, reducing to sine-Gordon in the real valued variable $\theta$.

\paragraph*{Evolution of the spectral transform.}

From the weak compatibility commutation \eqref{triplet} of operators $T_1$ and
$T_2$, the selected eigenfunctions $\psi^\pm$ are not eigenfunctions of $T_2$
and our purpose here is to determine their proper time evolution.  Equation
\eqref{triplet} implies the property
\begin{equation}
T_1\psi=0\ \Rightarrow\ (T_1-2\alpha_x)T_2\psi=0\ ,\end{equation}
and thus using the remarkable identity
\begin{equation}\label{prop}
T_1\psi=0\ \Rightarrow\ (T_1-2\alpha_x)(\psi_x+\alpha\psi)=0
\ ,\end{equation}
which simply results from the structure of $T_1$ in terms of the potential
$\alpha$, we may write 
\begin{equation}\label{base-evol}
T_2\psi^+=\Omega\ (\psi^+_x+\alpha\psi^+)\ .
\end{equation} 
Hereabove, the $x$-independent coefficient $\Omega(k,t)$ is called
the {\em generalized dispersion relation}. 
Note that one should actually expand $T_2\psi^+$ on a basis, say 
$\psi^+_x(k)+\alpha\psi^+(k)$ and  $\psi^+_x(-k)+\alpha\psi^+(-k)$, but
examination of the asymptotic behaviors as $x\to\infty$ readily reduces
the expansion to the form \eqref{base-evol} and gives  the evolution
\begin{equation}\label{evol-tau}
\tau_t=\left[\Omega-\frac1{4ik}\right]\tau\ ,
\end{equation}
of the transmission coefficient.

The value of the dispersion relation $\Omega$ together with the time 
evolution of $\rho$ are obtained by expressing 
equation \eqref{base-evol} and its $x$-derivative in $x=0$. We obtain
\begin{align}
(\alpha_0-ik)\rho_t+&\beta_0(1+\rho)=\Omega[\alpha_0(1+\rho)+ik(1-\rho)]\ ,\\
(\alpha_0^2-ik\alpha_0-k^2)\rho_t+&2\alpha_0\beta_0(1+\rho)
+ik\beta_0(1-\rho)=\notag\\
&\Omega[(\alpha_0^2-k^2)(1+\rho)+ik\alpha_0(1-\rho)]\ .
\end{align}
where we have used the expressions $ p=\alpha^2-\alpha_x$ and
$\beta_x=-2\alpha\beta$ and the equation $T_1\psi=0$.
This is a system of equations for the unknowns $\{\rho_t,\ \Omega\}$ in terms
of the boundary data $\alpha_0$ and $\beta_0$ and of $\rho$.
Note that it is compatible with (see appendix \ref{app:k=0})
\begin{equation}
k=0\ \Rightarrow\ \{\rho=-1,\ \tau=0\}\ .\end{equation}
Multiplying the first equation by $\alpha_0$ and substracting with the second
gives
\begin{equation}
\rho_t=(1+\rho)\left[\Omega+\beta_0\frac{\alpha_0}{k^2}\right]-
\frac{\beta_0}{ik}(1-\rho)\ .\end{equation}
Replaced in any of the preceding equations, we obtain the dispersion relation
which after some algebra can be written
\begin{equation}\label{Omega}
\Omega=\rho\ \frac{\alpha_0\beta_0}{2ik^3}(\alpha_0-2ik)
+\frac{\beta_0}{2ik^3}(\alpha_0^2+2k^2)\ .
\end{equation}
Thus the evolution of the reflection coefficient eventually reads as the
Riccati differential equation
\begin{equation}\label{evol-rho}
\rho_t=a\rho^2+b\rho+c\ ,
\end{equation}
with entries which depend on the boundary data by
\begin{equation}
a=\frac{\alpha_0\beta_0}{2ik^3}\ (\alpha_0-2ik)\ ,\quad 
b=\frac{\beta_0}{ik^3}\ (\alpha_0^2+2k^2)
\ ,\quad c=\frac{\alpha_0\beta_0}{2ik^3}\ (\alpha_0+2ik)
\ .\end{equation}

\paragraph*{Soliton generation.}

To illustrate the above formalism we consider the case of a piecewise constant
boundary datum, i.e.
\begin{equation}t=0\ :\ \tilde\theta(x)=0\ ,\quad
t<t_1\ :\ \theta_0(t)=0\ ,\quad t>t_1\ :\ \theta_0(t)=\varphi\ .
\end{equation}
For such a simple case, the Zakharov-Shabat spectral transform has been
shown to generate an infinite number of discrete eigenvalues with $k=0$ as
an accumulation point \cite{alex}.

Here we have from \eqref{alpha-0}
\begin{align}
&t<t_1\ :\ \alpha_0(t)=0,\ \beta_0=\frac14\ ,\notag\\
&t>t_1\ :\ \alpha_0(t)=\frac i2(t-t_1)\sin\varphi,\ \beta_0=\frac14e^{i\varphi}
\ .\end{align}
For such expressions, the time-evolution \eqref{evol-rho}, with vanishing 
initial condition $\rho(k,0)=0$, cannot be explicitely solved. However we can 
evaluate the large time asymptotic behavior as indeed
\begin{align}
&\frac ba=2\frac{\alpha_0^2+2k^2}{\alpha_0(\alpha_0-2ik)}
\underset{t\to\infty}\longrightarrow  2\ ,\notag\\
&\frac ca=\frac{\alpha_0+2ik}{\alpha_0-2ik}\underset{t\to\infty}
\longrightarrow 1\ .\end{align}
The resulting asymptotic evolution then reads
\begin{equation}
\rho_t\sim a(\rho+1)^2\ ,\end{equation}
which is easily solved to eventually get for large time $t$
\begin{equation}
\rho(k,t)\sim 
\frac{-e^{i\varphi}\sin\varphi[\frac13(t-t_1)\sin\varphi-k]\ (t-t_1)^2}
{32ik^3+e^{i\varphi}\sin\varphi[\frac13(t-t_1)\sin\varphi-k]\ (t-t_1)^2}\ .
\end{equation}
This function has 3 simple poles $k_n(t)$ given by
\begin{equation}
k_n(t)=(t-t_1)Z_n\ ,\quad 32iZ_n^3-2e^{i\varphi}\sin\varphi\ Z_n+
\frac13e^{i\varphi}\sin^2\varphi= 0\ .\end{equation}
For different values of $\varphi$, one or two of these poles are in the upper 
half plane corresponding to a  solution with a one-soliton or a two-soliton 
component.

Note that the Riccati time evolution \eqref{evol-rho} is not in general
explicitely solvable. Indeed, simple explicit solution could only be obtained
for $b/a$ and $c/a$ time independent which would immediately lead to
$\theta_0(t)=0$. Such does not occur in the Zakharov-Shabat spectral transform
of sine-Gordon where piecewise constant boundary data in $x=0$ produce an
explicitely solvable evolution \cite{alex}. Then it has been shown that the
spectral transform contains an infinite series of poles, accordingly with
\cite{fokas}, see also \cite{pelloni}. The problem is that these poles possess
an accumulation point (the origin $k=0$) and hence some of the basic solutions
are not continuous in the vicinity of the origin. Using the present spectral
transform based on the Schr\"odinger spectral problem removes this problem and
thus provides the correct solution to the Dirichlet boundary value problem for
sine-Gordon in light-cone coordinates.

\section{Inverse problem\label{sec:invpro}}

Solving the inverse problem will consist in computing the potential $p(x)$ from
the datum of $\rho(k)$. This is done by solving first the Hilbert boundary
value problem \eqref{hilbert} which actually requires the knowledge of the
analytical properties of the functions $\phi^\pm(k)$ together with their
behaviors at large $k$.

\paragraph*{Analytical properties.}

The Volterra integral equation \eqref{phi-} allows to deduce that $\phi^-$
is an entire function of $k$ bounded in the lower half plane. The function
$\phi^+$ is not analytic but the function $\tilde\phi^+=\phi^+/\tau$
solves the Volterra integral equation
\begin{equation}
\tilde\phi^+=1-
\frac1{2ik}\int_x^{\infty}dy\ \tilde\phi^+(k,y)p(y)\left[1-e^{-2ik(x-y)}
\right]\ .\end{equation}
Hence $\phi^+/\tau$ is analytic in the upper half-plane and thus, from
definition \eqref{tau-def}, $1/\tau$ is also analytic. Consequently $\phi^+$
has a pole where $1/\tau$ has a zero, these are the {\em discrete eigenvalues}
of the spectral transform named $k_n$.

The definition \eqref{rho-def} of the reflection coefficient shows that $\rho$
can be extended to the upper half plane (unlike in the infinite line case where
the integral would run on the whole real axis) where it possess  the discrete
eigenvalues as poles. The integral equation \eqref{phi+} can be rewritten as
\begin{equation}
\phi^+=1+\rho e^{-2ikx}+
\frac1{2ik}\int_{0}^xdy\ \phi^+(k,y)p(y)[1-e^{-2ik(x-y)}]\ .\end{equation}
By taking the residue of the above in $k_n$ we get
\begin{equation}
e^{2ik_nx}\underset{k_n}{\rm Res}(\phi^+)=\rho_n -
\frac1{2ik_n}\int_{0}^x[e^{2ik_ny}\underset{k_n}{\rm Res}(\phi^+)]
 p(y)[1-e^{2ik_n(x-y)}]\ ,\end{equation}
where obviously
\begin{equation}
\rho_n=\underset{k_n}{\rm Res}(\rho)\ .\end{equation}
By comparaison of the above integral equation with the one for the function
$\phi^-(-k,x)$ read in $k=k_n$ we immediately conclude that
\begin{equation}\label{residue}
e^{2ik_nx}\underset{k_n}{\rm Res}(\phi^+(k,x))=\rho_n\phi^-(-k_n,x)
\ .\end{equation}
This closes the analytical properties needed to the solution of the Hilbert
boundary value problem. 

\paragraph*{Asymptotic behaviors.}

It is necessary now to determine the large $k$ behaviors of the solutions
$\phi^\pm$. By repeated integration by part, the solutions obey 
\begin{align}
&\phi^+(k,x)\sim 1+\frac1{2ik}\phi^{(1)}+\frac1{(2ik)^2}\phi^{(2)}+\cdots
\label{large-phi+}\\
&\phi^-(k,x)\sim 1+\frac1{2ik}\phi^{(1)}+\frac1{(2ik)^2}\left[\phi^{(2)}+
\varphi^{(2)}e^{-2ikx}\right]+\cdots\label{large-phi-}
\end{align}
with the following coefficients
\begin{align}
&\phi^{(1)}=\int_0^xdy\ p(y)\ ,\label{phi-1}\\
&\phi^{(2)}=-p(x)+\int_0^xdy\ p(y)\int_0^ydz\ p(z)\ ,\\
&\varphi^{(2)}=p_0\ .
\end{align}
The solution of the Hilbert problem will have to verify those behaviors,
with in particular the essential singularity for $\phi^-$ on the real axis.

The relation \eqref{phi-1} furnishes the means to obtain the
potential $p$ from the solution $\phi^+$ namely
\begin{equation}\label{p-phi1}
p(x)=\partial_x\phi^{(1)}(x)\ ,\end{equation}
where $\phi^{(1)}$ is the coefficient of $1/2ik$ in the large $k$ expansion
of $\phi^+(k,x)$. Note finally that the spectral transform $\rho(k)$ obeys
in the upper half-plane
\begin{equation}\label{asymp-rho}
\rho(k)=-\frac{p_0}{(2ik)^2}+{\cal O}(1/k^3)\ ,\end{equation}
obtained again by integration by part on the definition \eqref{rho-def}
with help of the behavior \eqref{large-phi+}. Note that this is consistent with
the boundary value of $\phi^+$ in $x=0$ given in \eqref{phi+zero} and the
large $k$ behavior \eqref{large-phi+}. Indeed
\begin{align*}
&\phi^+\underset{x\to0}{\longrightarrow}1+\rho
\underset{k\to\infty}{\longrightarrow}1-\frac{p_0}{(2ik)^2}+\cdots\\
&\phi^+\underset{k\to\infty}{\longrightarrow} 1+\frac1{2ik}\phi^{(1)}+
\frac1{(2ik)^2}\phi^{(2)}+\cdots
\underset{x\to0}{\longrightarrow}1-\frac{p_0}{(2ik)^2}+\cdots
\end{align*}
This  works as  well for $\phi^-$, hence in summary, the limits $k\to\infty$ 
and $x\to0$  commute.

\paragraph*{Solution of the Hilbert problem.}

The solution $\phi^\pm$ of the Hilbert boundary value problem \eqref{hilbert}
which verifies the preceding analytical properties and large $k$ behaviors
is given from the data of $\rho(k)$ in ${\rm Im}(k)\ge0$ (including
eigenvalues $k_n$ in the upper half-plane and related residues $\rho_n$) by
the Cauchy-Green {\em integral equation}
\begin{align}
\phi^-(k,x)=1&+\frac{1}{2i\pi}\int_{-\infty+i0}^{+\infty+i0}
\frac{d\lambda}{\lambda-k}\ 
\phi^-(-\lambda,x)\rho(\lambda)e^{-2i\lambda x}\notag\\
&+\sum_{n=1}^N\frac{\rho_ne^{-2ik_nx}}{k_n-k}\ 
\phi^-(-k_n,x)\ ,\quad {\rm Im}(k)\le 0\ ,\label{cauchy-}
\end{align}
and the {\em explicit formula}
\begin{align}
\phi^+(k,x)=1&+\frac{1}{2i\pi}\int_{-\infty-i0}^{+\infty-i0}
\frac{d\lambda}{\lambda-k}\ 
\phi^-(-\lambda,x)\rho(\lambda)e^{-2i\lambda x}\notag\\
&+\sum_{n=1}^N\frac{\rho_ne^{-2ik_nx}}{k_n-k}\ 
\phi^-(-k_n,x)\ ,\quad {\rm Im}(k)\ge 0\ ,\label{cauchy+}
\ .\end{align}
As shown for the first time in \cite{alex}, the choice of the particular
contour in the complex plane for the above expressions is essential.

Indeed, while  the factor $e^{-2i\lambda x}$ produces a bounded kernel for
$x>0$ in the lower half $\lambda$-plane, it does not so in the upper half
$\lambda$-plane, that is for the expression of $\phi^-$. Assuming no pole for
$\rho$ on the real axis, we can use contour integration on \eqref{cauchy-} to
write
\begin{align}
\phi^-(k,x)&=1+\frac{1}{2i\pi}\int_{-\infty-i0}^{+\infty-i0}
\frac{d\lambda}{\lambda-k}\ 
\phi^-(-\lambda,x)\rho(\lambda) e^{-2i\lambda x}\notag\\
&-\phi^-(-k,x)\rho(k) e^{-2ikx}
+\sum_{n=1}^N\frac{\rho_ne^{-2ik_nx}}{k_n-k}\ 
\phi^-(-k_n,x) .\label{cauchy-bis}
\end{align}
The first consequence is that indeed these solution obey the Hilbert relation
\eqref{hilbert}. The second consequence is that they do verify the asymptotic
behaviors \eqref{large-phi+} and \eqref{large-phi-} obtained here simply by 
expanding the ratio $1/(\lambda-k)$ in expressions \eqref{cauchy+} and 
\eqref{cauchy-bis} where kernels are bounded, unlike in \eqref{cauchy-}.

The analytical properties naturaly follow the expressions \eqref{cauchy+} and
\eqref{cauchy-}. Indeed, as ${\rm Im}(k_n)>0$, the analyticity of the solution
$\phi^-(k,x)$ of \eqref{cauchy-} simply result from the property (continuity,
boundedness) of $\rho(k)$ on the real axis. Then the solution $\phi^+(k,x)$,
given by the explicit expression \eqref{cauchy+}, appears as the sum of an
analytic function and the contribution of its simple poles $k_n$.  Finally the
relation \eqref{residue} between the residue of $\phi^+$ in $k_n$ and $\rho_n$
readily follows from expression \eqref{cauchy+}.

\paragraph*{Reconstruction of $\theta(x,t)$.}

Once the spectral transform $\rho(k,t)$ has been obtained by solving the
Riccati time evolution \eqref{evol-rho} for the initial datum $\rho(k,0)$, the
solution of sine-Gordon $\theta(x,t)$ is built out of the integral equation
\eqref{cauchy-} as follows. The relation \eqref{p-phi1} gives the value of
$p(x,t)$ from the coefficient $\phi^{(1)}(x,t)$ of $1/2ik$ in the large $k$
expansion of $\phi^+(k,x,t)$, namely
\begin{align}\label{phi1}
\phi^{(1)}(x,t)=&-\frac{1}{\pi}\int_{-\infty-i0}^{+\infty-i0}d\lambda\ 
\phi^-(-\lambda,x,t)\rho(\lambda,t)e^{-2i\lambda x}\notag\\
&-2i\sum_{n=1}^N\phi^-(-k_n,x,t)\rho_ne^{-2ik_nx}\ .\end{align}
This expression shows the two building components of the solution: 
the radiation component represented by the continuous integral and the soliton
component represented by the discrete sum. Note that the number $N$ and 
positions $k_n$ of the solitons depend on time as allowed by the Riccati
time evolution \eqref{evol-rho}.

The simplest way to obtain now the solution $\theta(x,t)$ is to use the 
relations
\begin{equation} 
p_t=-2\beta_x\ ,\quad p=\phi^{(1)}_x\ ,\end{equation}
easily integrated with the values in $x=0$ of $\beta$. We eventually arrive at
\begin{equation}
e^{i\theta(x,t)}=e^{i\theta_0(t)}-2\partial_t\phi^{(1)}(x,t)\ .\end{equation}
This provides then  the solution $\theta(x,t)$ from 
the boundary value $\theta_0(t)$ and the solution $\phi^-(k,x,t)$ of the
integral equation \eqref{cauchy-} completely determined by $\rho(k,t)$.

\section{Conclusion}

The solution of the Dirichlet boundary value problem for the  sine-Gordon
equation has been shown to be completely characterized by the time-evolution
\eqref{evol-rho} of the spectral transform $\rho(k,t)$ evolving explicitely the
boundary datum $\theta(0,t)$, with an initial datum $\rho(k,0)$ determined from
the initial condition $\theta(x,0)$.

This solution has many important aspects. First the solution is {\em stable}
for initial conditions in the Schwartz space, much like what occurs for the 
infinite line case. Second, the time evolution of the spectral transform
is {\em regular} in the sense that it does not produces accumulation of poles.
Last but not least this approach allows to treat the {\em finite line} case as
indeed it is an open-end boundary value problem. This property was already
remarked in \cite{sasha} and later discussed in \cite{jl-boit}, and mainly
results from the fact that the solution is completely determined by the
integral equation \eqref{cauchy-} where the $x$-dependence is {\em parametric}.

In order now to descibe in more details interesting generic behaviors of the
solution of sine-Gordon, we would need to study the Riccati time evolution
\eqref{evol-rho} where the coefficients are time-dependent. This aspect of the
work is reported to further studies as it would require use of numerical
simulations.

\appendix

\section{Behaviors at the origin\label{app:k=0}}

The definition \eqref{phi+} of the solution $\phi^+(k,x)$ and the Hilbert
relation \eqref{hilbert} written both  in $x=0$ lead to the system
(in $k=0$ we note $\phi(0,x)=\phi_0(x)$ and $\rho(0)=\rho_0$)
\begin{align}
&\phi^+_0(x)=1+\rho_0-\int_x^\infty dy(x-y)\phi^+_0(y)p(y)\ ,
\label{phi0+}\\
&\phi_0^+(x)=[1+\rho_0]\phi_0^-(x)\ ,\label{hilb0}
\end{align}
while $\phi_0^-$ solves
\begin{equation}\label{phi0-}
\phi_0^-(x)=1+\int_0^x dy(x-y)\phi_0^-(y)p(y)\ .\end{equation}
As the above Volterra integral equation has a unique solution, $\phi_0^-$
does not vanish and the equation \eqref{phi0+} can be written by using
\eqref{hilb0} as
\begin{equation}\label{phi0+bis}
\phi_0^+(x)\left[1-1/\phi_0^-(x)\right]=-\int_x^\infty dy(x-y)\phi^+_0(y)p(y)\ .
\end{equation}
This has the unique vanishing solution and consequently we have proved that
\begin{equation}
\phi_0^+(x)=0\ ,\quad \rho_0=-1\ .\end{equation}
Finally the unitarity relation \eqref{unitarity} leads immediately 
$\tau(0)=0$ as expected.

\section{Reduction}

We determine here the symmetry properties of the
basic solutions and the spectral transform resulting from the constraint
$\theta\in\mathbb R$. First we note that
\begin{equation}\label{T1-star}
\theta\in{\mathbb R}\ \Rightarrow\ T_1^*=T_1-2\alpha_x\ ,\end{equation}
where for any function $f(k)$ of the complex variable $k$, we define
\begin{equation}f^*=\overline f(\overline k)\ .\end{equation}
Second the operator $T_1$ has the property \eqref{prop} and consequently when
$\theta$ is real valued (or equivalently when $\alpha$ is purely imaginary),
the eigenfunctions $\psi^*$ and $\psi_x+\alpha\psi$ are linearly dependent,
namely
\begin{align}
&(\partial_x+\alpha)\psi^+(k)=A\overline\psi^+(\overline k)+
B\overline\psi^+(-\overline k)\ ,\label{exp-reduc+}\\
&(\partial_x+\alpha)\psi^-(k)=C\overline\psi^+(\overline k)+
D\overline\psi^+(-\overline k)\ .\label{exp-reduc-}\end{align}
The coefficients hereabove are computed by expressing these expansions in
$x=\infty$ with help of \eqref{phi+infty} and \eqref{phi-infty}. We obtain
using $\alpha(\infty)=0$
\begin{align}
& A=0  \ , & B=ik \frac{\tau(k)}{\overline\tau(-\overline k)}\ ,\\
& C=ik\frac{\rho(k)}{\tau(k)\overline\tau(\overline k)} 
\ , & D=\frac{ik}{\tau(-k)\overline\tau(-\overline k)}\ .
\end{align}

The next step consists in expressing \eqref{exp-reduc+} and \eqref{exp-reduc-}
in $x=0$ which results in a set of two equations linking the spectral
coefficients and the boundary value $\alpha_0$.  By eliminating $\alpha_0$
between these two equations and after some simple algebra using the unitarity
relation \eqref{unitarity} we eventually obtain first
\begin{equation}\label{reduction}
\frac12\frac{[1+\overline\rho(\overline k)][1+\rho(k)]}
{\tau(k)\overline\tau(\overline k)}+
\frac12\frac{[1+\overline\rho(-\overline k)][1+\rho(-k)]}
{\tau(-k)\overline\tau(-\overline k)}=1\ ,\end{equation}
which constitutes the condition for a spectral transform $\rho(k)$
to ensure a {\em real valued} reconstructed field $\theta(x)$.
The second equation gives a relation between the boundary value $\alpha_0$
and the spectral transform, namely
\begin{equation}\label{alpha-rho}
\alpha_0=-ik\left[1-
\frac{1+\overline\rho(-\overline k)}{\tau(-k)\overline\tau(-\overline k)}
-\rho(k)\frac{1+\overline\rho(\overline k)}{\tau(k)\overline\tau(\overline k)}
\right]\ .\end{equation}

This relation must in particular be valid as $k\to\infty$ where the behaviors
of the spectral coefficients $\rho$ and $\tau$ are given by
\begin{align}
&\rho(k)\sim\frac{p_0}{4k^2}+{\cal O}(\frac1{k^3})\ ,\\
&\tau(k)\sim 1+\frac\gamma{2ik}+
\frac{\gamma^2}{2(2ik)^2}+{\cal O}(\frac1{k^3})\ .
\end{align}
The behavior of $\tau$ is obtained from its definition \eqref{tau-def}
and the large $k$-behavior \eqref{large-phi+} of the eigenfunction $\phi^+$.
By using then the relation \eqref{phi-1} it is not difficult to obtain
\begin{equation}
\gamma=\alpha_0+\int_0^\infty dx\ \alpha^2\ .\end{equation}
The reduction constraint $\theta\in\mathbb R$  is here equivalent to 
$\gamma-\overline\gamma=2\alpha_0$. With all this in hands, the above
relation \eqref{alpha-rho} does hold as $k\to\infty$ where the zeroth order
term in the right hand side is indeed $\alpha_0$. One can check then
that the next order is a vanishing identity as we have within the reduction
\begin{equation}
p_0+\overline p_0-2\alpha_0^2=0\ .\end{equation}


\begin{thebibliography}{aaaaaa}
\bibitem{ist-sg} M.J. Ablowitz, D.J. Kaup, A.C. Newell, H. Segur, 
Phys Rev Lett {\bf 30} (1973) 1262. 
\bibitem{takh-fad}L.A. Takhtajan, L.D. Faddeev,  Theor Math Phys {\bf 21} 
(1974) 1046
\bibitem{fokas}A.S. Fokas, Proc Roy Soc Lond A {\bf 453} (1997) 1411
\bibitem{alex}J. Leon, A. Spire, J Phys A {\bf 34} (2001) 7359
\bibitem{marflo}M. Boiti, J. Leon, F. Pempinelli, Inv Problems {\bf 3} (1987) 37
\bibitem{sasha} J. Leon, A.V. Mikhailov,  Phys Lett A {\bf 253} (1999) 33
\bibitem{jl-fx} F-X. Hugot, J. Leon, Inv Problems {\bf 15} (1999) 701
\bibitem{jl-boit} M. Boiti, J. Leon, F. Pempinelli, J Phys A {\bf 32} (1999) 927
\bibitem{pierre} P.C. Sabatier, J Math Phys, {\bf 41} (2000) 414
\bibitem{flomar}M. Boiti, J. Leon, M. Manna, F. Pempinelli,
Inv Problems {\bf 2} (1986) 271; {\em ibid} {\bf 3} (1987) 25
\bibitem{thana-bea}A.S. Fokas, B. Pelloni, 
Proc Roy Soc Lond A {\bf 456} (2000) 805
\bibitem{sabatier}K.H. Chadan, P.C. Sabatier, {\em Inverse problems in quantum 
scattering theory}, Springer Verlag (New-York 1989). B.M. Levitan, 
I.S. Sargsjan, {\em Introduction to spectral theory}, 
Transl Math Monographs, A.M.S. (Providence 1975).
\bibitem{pelloni}B. Pelloni, Comm Appl Anal {\bf 6} (2002) 179
\end{thebibliography}
\end{document}